\documentclass[aps, prd, groupedaddress, showkeys,twocolumn]{revtex4}
\usepackage{bm,amsmath}
\usepackage{amsthm,amssymb}
\usepackage{graphicx}
\usepackage{dcolumn}
\usepackage{mathrsfs}
\usepackage{amsmath}
\usepackage{latexsym}
\usepackage{multirow}
\def\eq#1{(\ref{#1})}

\def\al{\alpha} 
\def\be{\beta} 
\def\ga{\gamma}

\def\ep{\epsilon}

\def\et{\eta}

\def\ka{\kappa}
\def\la{\lambda}

\def\si{\sigma}

\def\La{\Lambda}

\def\Om{\Omega}

\newcommand{\ben}{\begin{equation}}
\newcommand{\een}{\end{equation}}
\newcommand{\bea}{\begin{eqnarray}}
\newcommand{\eea}{\end{eqnarray}}
\newcommand{\ba}{\begin{array}}
\newcommand{\ea}{\end{array}}
\newcommand{\bit}{\begin{itemize}}
\newcommand{\eit}{\end{itemize}}

\def\math{\mathsurround 0pt}
\def\oversim#1#2{\lower.5pt\vbox{\baselineskip0pt \lineskip-.5pt
        \ialign{$\math#1\hfil##\hfil$\crcr#2\crcr{\scriptstyle\sim}\crcr}}}

\def\pa{\partial}

\def\half{\frac{1}{2}}

\newcommand{\etRG}{{\et_{\rm RG}}}

\newcommand{\nuRG}{{\nu_{\rm RG}}}
\newcommand{\alRG}{{\al_{\rm RG}}}
\newcommand{\siRG}{{\si_{\rm RG}}}

\begin{document}

\title{Asymptotically Safe Cosmology}
\author{Mark Hindmarsh}
\author{Daniel Litim}
\author{Christoph Rahmede}
\affiliation{Department of Physics \& Astronomy, 
University of Sussex, Brighton BN1 9QH, UK}

\begin{abstract}
We study quantum modifications to cosmology  in a Friedmann-Robertson-Walker universe with and without scalar fields by taking the renormalisation group running of gravitational and matter couplings into account. We exploit the Bianchi identity to relate the renormalisation group scale with scale factor and derive the improved cosmological evolution equations.
We find two types of cosmological fixed points where the renormalisation group scale either freezes in, or continues to evolve with scale factor. We discuss the implications of each of these, and classify the different cosmological fixed points with and without gravity displaying an asymptotically safe renormalisation group fixed point. We state conditions of existence for an inflating ultraviolet cosmological fixed point for Einstein gravity coupled to a scalar field. We also discuss other fixed point solutions such as ``scaling"  solutions,  or fixed points with equipartition between kinetic and potential energies. 
\end{abstract}

\pacs{Pacs numbers}

\maketitle

\section{Introduction}

Quantum gravity effects become important at very early times in the history of the Universe, when the Planck scale is approached. They add to the already present quantum fluctuations in the  dynamics of matter fields. 
Provided that the metric field remains the fundamental carrier of the gravitational force,  quantum corrections should take the form of a simple modification of  couplings whose scale-dependence is described by the Renormalisation Group (RG). 
For gravity, such a scenario would imply that the ultraviolet behaviour of the gravitational couplings is dictated by a renormalisation group fixed point \cite{Weinberg:1980gg}.\medskip

In the past decade,  good evidence for this asymptotic safety scenario has been achieved
based on RG studies in the continuum  \cite{Litim:2006dx,Niedermaier:2006wt,Niedermaier:2006ns,Percacci:2007sz,Litim:2008tt}
and numerical studies on the lattice \cite{Ambjorn:2001cv,
Hamber:1999nu,Hamber:2004ew}. This has lead to the exciting idea that an asymptotically safe UV fixed point may control the 
beginning of the Universe \cite{Bonanno:2001xi}, an idea which subsequently  has been explored in Einstein gravity  with an ideal fluid 
\cite{Bonanno:2001xi,Bonanno:2002zb,Koch:2010nn}, and in the context of $f(R)$ gravity \cite{Weinberg:2009wa, Tye:2010an,Bonanno:2010bt,Contillo:2010ju}. There is also the possibility of an infrared (IR) fixed point which can play a role in the observed acceleration of the Universe today \cite{Bonanno:2001hi,Bentivegna:2003rr,Reuter:2005kb,Babic:2004ev}.  Perturbative RG approaches to cosmology have  explored the running of the cosmological constant~ \cite{Shapiro:1999zt,Shapiro:2000dz,Shapiro:2008yu,Shapiro:2009dh} and of Newton's coupling \cite{Shapiro:2004ch,Bauer:2005rpa,Bauer:2005tu} and their implications for big bang nucleosynthesis, supernova observations, and deviations from standard cosmology \cite{Guberina:2002wt,Shapiro:2003kv,EspanaBonet:2003vk,Shapiro:2004ch}.
\medskip

An important question in the above is how to relate the RG scale parameter to cosmological time. Reuter and collaborators~\cite{Bonanno:2001xi,Bonanno:2001hi,Reuter:2005kb} have chosen the RG scale inversely proportional to cosmological time, also exploring a matching with the Hubble scale. Other approaches link the RG scale with eg.~the fourth root of the energy density \cite{Guberina:2002wt}, or the cosmological event and particle horizons \cite{Bauer:2005rpa,Bauer:2005tu}. 
This leads to RG improved cosmological equations, which may even generate entropy \cite{Bonanno:2007wg,Bonanno:2008xp}.  \medskip

Our work is motivated by 
the fundamental role played by scalar fields in cosmology,  
which generalise a cosmological constant.
It is therefore natural to adapt cosmological RG ideas 
to include a scalar field. 
Our viewpoint is that we wish to choose the RG scale as to maintain the form of the classical equations.
Taking into account the scale-dependence of the scalar field couplings, the preservation of the Bianchi identity  
leads to an evolution equation for the RG scale rather than an algebraic condition as hitherto found.
 Under some conditions, which we detail, it is true that the RG scale evolves inversely proportional to cosmic time (or more precisely, the Hubble parameter $H$), but this behaviour is not found in general.\medskip

In this paper, we implement the above ideas by formulating 
 the cosmological  evolution equations 
as an autonomous system evolving with the scale factor \cite{Copeland:1997et}.  In analogy with the RG evolution, these cosmological equations have fixed points, which may be approached with decreasing or increasing scale factor.  We find that
the universe can asymptote to a simultaneous fixed point of both the RG and cosmological evolution equations at large and small scale factors.
In addition, we find new cosmological fixed points where the RG scale freezes at some constant value.
We exhibit several of these, showing how they relate to fixed points of the classical evolution.
Whether there is scalar-field driven inflation at these fixed points depends on the parameters describing the RG running of the gravitational couplings and the scalar potential. 

\section{Cosmology}
\label{sec:infla}

In this section, we recall the standard set-up and introduce our notation and conventions. We assume that the universe is described by Einstein gravity with a spatially flat Friedmann-Robertson-Walker (FRW) metric, and contains 
a minimally coupled scalar field with arbitrary potential $V(\phi)$ and  
fluids with barotropic equations of state $p_i=w_i \rho_i$.  We assume that in the periods of interest the energy-momentum is  separately conserved. 
Defining $\gamma_i = 1+w_i$, the dynamical evolution of the system is controlled by the Einstein equations and the scalar field equation of motion, which 
under our assumptions give
\begin{eqnarray}
\dot{H}&=&-\frac{\kappa^2}{2}\left(\sum_i\ga_i\rho_i+\dot{\phi}^2\right) \label{eq:Hdot}\\
\dot{\rho}_i&=&-3H \ga_i \rho_i \label{eq:rhodot}\\
\ddot{\phi}&=&-3H\dot{\phi}-\frac{{d}V}{{d}\phi} \label{eq:phiddot}
\end{eqnarray}
together with the Friedmann constraint
\begin{equation}\label{Friedmann}
H^2=\frac{\kappa^2}{3}\left(\sum_i\rho_i+\frac{\dot{\phi}^2}{2}+V(\phi)\right). 
\end{equation}
Here $\kappa^2=8\pi G$, $G$ is Newton's constant, $H=\dot{a}/a$ is the Hubble parameter, and the energy density of a homogeneous scalar field is $\rho_{\phi}=\frac 12\dot{\phi}^2+V(\phi)$.
We view a cosmological constant $\Lambda$ as being contained in the potential as $V(\phi)|_{\phi_{\rm min}}=\Lambda/(8\pi G)$.\medskip

Following \cite{Copeland:1997et,Copeland:2006wr} we introduce the dimensionless dynamical variables 
\ben\label{xyz}
x = \frac{\kappa\dot{\phi}}{\sqrt{6}H}, \ \ y=\frac{\kappa\sqrt{V}}{\sqrt{3}H}, \ \  z = \frac{V'}{\kappa V}\ .
\een
Note that $x$ and $z$ are related to the traditional slow-roll parameters by $\ep_H = 3x^2$ and $\ep_V =\frac 12 z^2$, where 
$\ep_H = - \dot H/(H^2)$ and $\ep_V = \frac 12 (\ka V'/V)^2$, where $\Omega_i=0$ as in traditional cold inflation.
We also define $N=-\ln a$ (note the sign convention) 
and denote 
the density parameter of the $i$th barotropic fluid as $\Om_i$.
The equations of motion can then be rewritten in  autonomous form
\begin{eqnarray} 
\frac{{d}x}{{d}N}&=&
3x(1-x^2)+\sqrt{\frac32}y^2z -\frac{3}{2}x\sum_i\ga_i\Om_i\ ,\\
\frac{{d}y}{{d}N}&=&
-\sqrt{\frac32}x yz - 3 x^2 y - \frac{3}{2}y\sum_i\ga_i\Om_i\ ,\\
\frac{{d}z}{{d}N}&=&-\sqrt{6}\,x(\eta-z^2)\ , \label{eq:ZeqnClassical}\\
\frac{d\Om_i}{dN}&=&-3\Om_i\left( 2x^2+\sum_j\ga_j\Om_j - \ga_i \right)\,,
\end{eqnarray}
with the Friedmann constraint expressed as
\ben\label{eq:FriCon}
1 - x^2 - y^2 - \sum_i\Om_i=0.
\een
The new quantity $\eta = V''/(\ka^2 V)$ is to be viewed as a function of $z$, which requires inverting the equation $z = V'(\phi)/(\ka V(\phi))$. For example, potentials which are field monomials $V = \la_n\phi^n$ give $\eta(z) = (n-1)z^2/n$. For more complex potentials $\et(z)$ is not generally expressible in closed form. \medskip

The dependence of $H$ on scale factor is given by
\ben\label{dHdN}
\frac{d \ln H}{dN} = 3x^2 + \frac{3}{2}\sum\ga_i\Om_i\ .
\een
In our conventions, the condition for inflation reads $d\ln H/d N<1$.\medskip

The equations of motion take the form of a flow with a constraint.  The flow of the system can have fixed points, where 
\begin{equation}\label{FP}
\frac{d}{dN}(x,y,z,\Omega_i)=0
\end{equation}
which represent the allowed states from which a FRW universe can emerge in the past ($N\to\infty$) or evolve towards in the future ($N\to - \infty$).
To obtain the fixed points, a solution to the equation $\et = z^2$ must be found.  Denoting such a solution by $z_*$, we tabulate the fixed points for a scalar field and one other barotropic fluid in Tab.~\ref{table1}, reproducing the results of Copeland et al. \cite{Copeland:1997et}.
In this language, inflation can be regarded as an emergence from the potential-dominated fixed point (a).  Flows can leave this fixed point only along the line $x = -z/\sqrt 6$, which we recognise as the slow-roll equation for inflation.
Note that $\et = z^2$ with $x,z\ne 0$ can only be produced by potentials of the form $V = V_0\exp(\la\ka\phi)$, with $\la$ a dimensionless constant. The fact that $x\ne 0$ implies that $\phi$ is evolving at the cosmological fixed point, and $z\ne 0 $ implies that $z$ changes with $\phi$: therefore $\et = z^2$ must apply for a range of $z$, and is trivially integrable to find $V$.

\begin{table*}[t!]
\begin{center}
\begin{tabular}{|c|c|c|c|c|c|c|} \hline
{\multirow{3}{*}{Case}}
&{\multirow{3}{*}{$x$}} 
&{\multirow{3}{*}{$y$}}
 & {\multirow{3}{*}{$z$}}
 & {\multirow{3}{*}{$\Omega_{\gamma}$}} 
 &{\multirow{3}{*}{ existence}} 
&{\multirow{3}{*}{type}}
\\&&&&&&
\\&&&&&&
\\ \hline
{\multirow{3}{*}{(a)}}
&{\multirow{3}{*}{ 0}}
 &  {\multirow{3}{*}{$1$ }}
 &{\multirow{3}{*}{ 0}}
& {\multirow{3}{*}{ 0}}
 & {\multirow{3}{*}{all $\gamma$}} 
  &{\multirow{3}{*}{potential}}
  \\
  &&&&&&
  \\
  &&&&&&
\\ \hline
{\multirow{3}{*}{(b)}}
&{\multirow{3}{*}{  $ \pm 1$ }}
&{\multirow{3}{*}{  0}}
 &{\multirow{3}{*}{ $z_*$}}
  &{\multirow{3}{*}{ 0}} 
   &{\multirow{3}{*}{all $\gamma$ and  $z_*$}}
 &{\multirow{3}{*}{kinetic}}
 \\
 &&&&&&
 \\
  &&&&&&
\\ \hline
{\multirow{3}{*}{(c)}}
& {\multirow{3}{*}{   $\displaystyle - \frac{z_*}{\sqrt{6}}$ }}
&{\multirow{3}{*}{  $\displaystyle \sqrt{1 -
 \frac{z_*^2}{6}}$}} 
 &{\multirow{3}{*}{ $z_*$}}
 & {\multirow{3}{*}{0}}
 &{\multirow{3}{*}{ $z_*^2 \le6$}} 
&{\multirow{3}{*}{mixed}}
\\
&&&&&&
\\
 &&&&&&
   \\ \hline
{\multirow{3}{*}{(d)}}&
{\multirow{3}{*}{  $\displaystyle - \sqrt{\frac{3}{2}} \frac{\ga}{z_*}$ }}
&{\multirow{3}{*}{ $\displaystyle  
\sqrt{\frac{3}{2} \frac{\ga(2-\ga)}{z_*^2}}$}}
 &
 {\multirow{3}{*}{ $z_*$}}
 &{\multirow{3}{*}{ $\displaystyle 1 - \frac{3\ga}{z_*^2}$}}
&{\multirow{3}{*}{  $z_*^2 \ge 3\ga $}}  
 &{\multirow{3}{*}{scaling}}
 \\
 &&&&&&
 \\
 &&&&&&
\\ \hline
{\multirow{3}{*}{(e)}}
&{\multirow{3}{*}{0}}
&{\multirow{3}{*}{0}} 
& {\multirow{3}{*}{--}} 
& {\multirow{3}{*}{1}} 
&{\multirow{3}{*}{$\gamma\neq 0$}}
&{\multirow{3}{*}{fluid}}
\\
&&&&&&
\\
  &&&&&&
\\ \hline
\end{tabular}
\caption{\label{table1}{Classical cosmological fixed points for a scalar field
 and one other barotropic fluid with equation of state parameter 
 $\gamma=1+w$. In all cases, $z_{*}$ is a solution to the equation
  $\et(z) = z^2$, see \eq{eq:ZeqnClassical}. Fixed point (d) exists only for
$z_*^2 < 6$. Fixed point (e) exists only if $z_*^2 > 3\ga$.
For the case of an exponential potential, these fixed points and eigenvalues 
agree with Copeland et al. \cite{Copeland:1997et}. Note that for $z_*=0$  fixed point (a) is just a special case of (c). Fixed point (e) corresponds to the fluid dominated
solution, fixed points (a), (b), (c)  to scalar field dominated solutions with either
potential term (a) or kinetic term (b) dominating or a mixture (c), and fixed point (d) is known as the "scaling" or solution.}}
\end{center}

\end{table*}

\section{Renormalisation Group}    \label{RG}                                                                                                                                               

The cosmological equations of the previous section  derive from classical equations of motion, and thus rely on classical general relativity. In this section, we evaluate how quantum fluctuations of the metric field and the        
scalar field modify the cosmological dynamics. Once quantum fluctuations are taken into account, classical equations of motions as obtained from the classical action $S$ should be replaced by quantum equations of motions as obtained from the quantum effective action $\Gamma$. \medskip

A powerful method to obtain the quantum effective action is given by Wilson's renormalisation group, based on the successive integrating-out of momentum modes from a path-integral based formulation of field theory \cite{Berges:2000ew,Bagnuls:2000ae,Litim:2008tt}. This way, the  effective action becomes scale-dependent $\Gamma\to \Gamma_k$,  and interpolates, as a function of the RG momentum scale parameter $k$, between the fundamental action $S$  and the full quantum effective action $\Gamma$. The dependence of $\Gamma_k$ on the RG scale is given by an exact, functional differential equation \cite{Wetterich:1992yh}
\begin{equation}
\label{flow} k\partial_k\Gamma_k[\phi]=\frac12
{\rm Tr}\frac{1}{\Gamma^{(2)}_k[\Phi]+R_k}k\partial_k R_k\,, 
\end{equation} 
 which links the scale-dependence of
$\Gamma_k[\Phi]$ with its second functional derivative
$\Gamma_k^{(2)}[\Phi]\equiv\delta^2\Gamma_k
  [\Phi]/({\delta\Phi\delta\Phi})$. Here, $\Phi=(g_{\mu\nu}, \phi,\cdots)$ stands for all propagating fields in the theory. The trace denotes a momentum integration, and 
  $R_k(q)$  the Wilsonian infrared cutoff \cite{Berges:2000ew,Bagnuls:2000ae,Litim:2008tt}. For the RG scheme $R_k(q^2)\to k^2$, the flow \eq{flow} becomes a renormalised Callan-Symanzik equation. The RG scheme can be chosen to optimise the stability and convergence of the RG flow \cite{Litim:2000ci,Litim:2001up}.
The functional RG flow \eq{flow} can be seen as the generating functional for the RG equations for all couplings of the theory. Standard perturbation theory is reproduced when \eq{flow} is solved iteratively for small coupling \cite{Litim:2001ky,Litim:2002xm}. A particular strength of \eq{flow} is its applicability even for systems with strong correlations and  large couplings.\medskip

Explicit RG equations  for gravity, or gravity with matter fields, have been obtained with the help of Wilson's renormalisation group \eq{flow}, eg.~\cite{Reuter:1996cp,Dou:1997fg,Souma:1999at,Lauscher:2001ya,Lauscher:2002sq,Reuter:2001ag,Litim:2003vp,Percacci:2003jz,Percacci:2005wu,Litim:2006dx,Niedermaier:2006wt,Niedermaier:2006ns,Percacci:2007sz,Litim:2008tt,Codello:2006in,Codello:2007bd,Codello:2008vh,Narain:2009fy,Narain:2009gb,Benedetti:2009gn,Machado:2007ea}. 
For our purposes it is interesting to discuss the case where gravity and/or the matter sector display a non-trivial RG fixed point. For gravity, an ultraviolet (UV) fixed point implies that gravity becomes asymptotically safe at short distances 
 \cite{Weinberg:1980gg}. Substantial evidence for a gravitational fixed point has been generated in the past decade based on RG studies in the continuum \cite{Reuter:1996cp,Litim:2006dx,Niedermaier:2006wt,Niedermaier:2006ns,Percacci:2007sz,Litim:2008tt} and numerical simulations on the lattice \cite{Ambjorn:2001cv,Hamber:1999nu,Hamber:2004ew}. It has also been speculated that the long-distance behaviour of gravity might be described by an IR fixed point \cite{Bonanno:2001hi,Bentivegna:2003rr,Reuter:2005kb,Babic:2004ev}. To illustrate the role of a gravitational fixed point, we recall the RG equation for Newton's coupling, written in terms of the renormalised dimensionless coupling             
\begin{equation}                                                                                                                                                                  
\label{g}                                                                                                                                                                         
g(k)=G_k\,k^2\equiv G\,Z_N^{-1}(k)\,k^2\,.                                                                                                                                      
\end{equation}                                                                                                                                                                    
Here, $G$ denotes Newton's coupling, $G_k=G\,Z_N^{-1}(k)$ the renormalised coupling and  $Z_N(k)$ the graviton's wave function renormalisation. The RG             
equation for \eq{g} takes a simple form                                                                                                                                           
\begin{equation}                                                                                                                                                                  
\label{dg}                                                                                                                                                                        
\frac{dg}{d\ln k}=(2+\etRG)\,g(k)\,.
\end{equation}                                                                                                                                                                    
In general, the anomalous dimension is a function of all other couplings of the theory including Newton's coupling, the cosmological constant, and possible further gravity-matter couplings such as the one multiplying an interaction term $R\phi^2$, where $R$ is the Ricci scalar and $\phi$ a scalar field. 

In this paper we will take into account only Newton's coupling, the cosmological constant and the couplings in the scalar field potential as the main ingredients in standard early-universe cosmology. 
We shall ignore the running of the kinetic term through the scalar wave function factor, as it     
vanishes to one-loop order in scalar theories. 
Explicit expressions for the RG flows of the couplings have been obtained in \cite{Reuter:1996cp,Dou:1997fg,Souma:1999at,Lauscher:2001ya,Lauscher:2002sq,Reuter:2001ag,Litim:2003vp,Percacci:2003jz,Percacci:2005wu,Litim:2006dx,Niedermaier:2006wt,Niedermaier:2006ns,Percacci:2007sz,Litim:2008tt,Codello:2006in,Codello:2007bd,Codello:2008vh,Narain:2009fy,Narain:2009gb,Benedetti:2009gn,Machado:2007ea}. 
The resulting RG scale dependence $G\to G_k$, $\Lambda\to\Lambda_k$, and $V(\phi)\to V_k(\phi)$.
can be conveniently described in terms of the dimensionless parameters
\ben\label{eq:RGexpDef}                                                                                                                                                           
\etRG=\frac{\pa\ln G_k}{\pa\ln k}\,,\ \    \nuRG=\frac{\pa \ln V_k}{\pa \ln k}\,,\ \ \siRG = \frac{\pa \ln V'_k}{\pa \ln k}\ .                                                                                                   
\end{equation}        
Note that $\nuRG$ and $\siRG$ are also manifestly functions of the variable $z$ through     
their dependence on $\phi$. Equally, $\etRG$ can acquire a $z(\phi)$ dependence through radiative corrections. 
In our conventions the potential includes the cosmological constant $\Lambda_k$.
This implies that the RG dependence of the cosmological constant is expressed in terms of \eq{eq:RGexpDef} by
\ben\label{CC}
\frac{\pa \ln \La_k}{\pa \ln k} =\nuRG + \etRG\,.
\een                                                                          
In the classical regime, the anomalous dimension is very small           
$|\etRG|\ll 1$, and  \eq{dg} displays the Gaussian (free theory) fixed point $g_*=0$. In its vicinity, quantum effects are small and the gravitational couplings take their       
classical values $G_k\approx G$. However, once the anomalous dimension becomes large, \eq{dg} displays a non-trivial fixed point where  the anomalous dimension            
counter-balances the canonical dimension of Newton's coupling, $\etRG=-2$. In its vicinity, the renormalised Newton's coupling becomes very small (large), 
\begin{equation}\label{UV}
G_k\approx            
g_*/k^2\,,
\end{equation} 
provided it corresponds to an ultraviolet (infrared) fixed point. The fixed point scaling \eq{UV} leads to a characteristic weakening of gravity at shortest distances $G_k\ll G$. In the context of RG improved black hole solutions, this behaviour \eq{UV} increases the domain of validity of the semi-classical approximation \cite{Bonanno:2000ep,Falls:2010he}. The RG running of couplings towards the IR can also build up non-local corrections to the effective action, see \cite{ Satz:2010uu} for a recent example thereof. In our set-up, this is accounted for through the RG running of couplings.  Similarly, the effective potential $V_k$ achieves an RG fixed point provided it scales with its canonical dimension $V_k\propto k^4$, leading to $\nuRG=4$. This conclusion is equally reached by looking at the RG flow for the cosmological constant \eq{CC}: since the canonical dimension of $[\Lambda_k]=2$, a non-trivial RG  fixed point for the dimensionless coupling $\Lambda_k/k^2$ requires that $\nuRG+\etRG=2$.  \medskip

\section{RG improved cosmology}    \label{RGC}

In this section, we discuss how the classical cosmological equations  
are modified by quantum effects. 
In our framework, the structure of the classical dynamical equations 
is preserved in the sense that quantum corrections appear as $k$-dependences of the gravitational coupling $G_k$ and      
of the parameters in the scalar potential $V_k(\phi)$, in a functional form fixed by the RG equations \eq{eq:RGexpDef}. This turns the variables \eq{xyz} into scale-dependent variables. The RG parameters should be regarded as given functions of the RG scale $k$. It is the central physics assumption of this paper that the RG scale parameter is a function of cosmological time, $k=k(t)$, which we view in our conventions as $k=k(N)$. Consequently, 
time-dependence of the RG corrections is determined by the time evolution via $k(N)$.
We assume throughout that homogeneity and isotropy are maintained. 
The validity of the classical equations is justified by the            
anti-screening behaviour of gravity with decreasing $G$. Possible non-local     
contributions are effectively taken into account by making $k$                  
time-dependent as argued in \cite{Bonanno:2010bt} for the case $k \propto      
H$.  We will see that the Bianchi identity can force a different time           
dependence, but the general argument remains valid.
\medskip

Enhancing the classical equations \eq{eq:Hdot} -- \eq{Friedmann}  with the help of the RG equations \eq{eq:RGexpDef}, and also using the definition \eq{xyz} with couplings replaced by running couplings, we arrive at the RG improved cosmological equations
\begin{widetext}
\begin{eqnarray} 
\frac{{d}x}{{d}N}&=&
3x(1-x^2)+\sqrt{\frac32}y^2z -\frac{3}{2}x\sum_i\ga_i\Om_i  + \half x \etRG  \frac{d \ln k}{dN} , \label{eq:XeqnRG}\\
\frac{{d}y}{{d}N}&=&
-\sqrt{\frac32}x yz - 3 x^2 y - \frac{3}{2}y\sum_i\ga_i\Om_i +  \half y (\etRG + \nuRG)  \frac{d \ln k}{dN} \ ,\label{eq:YeqnRG}\\
\frac{{d}z}{{d}N}&=&-\sqrt{6}\,x(\eta-z^2) + z\left(-\half \etRG - \nuRG + \siRG\right)  \frac{d \ln k}{dN}\ , \label{eq:ZeqnRG}\\
\frac{d\Om_i}{dN}&=&-3\Om_i\left( 2x^2+\sum_j\ga_j\Om_j - \ga_i \right)  + \etRG\Om_i \frac{d \ln k}{dN}\ . \label{eq:OeqnRG}
\end{eqnarray}
\end{widetext}
These dynamical equations are subject to the Friedmann constraint, and an extra constraint deriving from the Bianchi identity. Our working hypothesis -- which is that the functional form of the dynamical equations is preserved -- implies that the energy-momentum tensor remains covariantly conserved for any fixed $k$. Hence the Bianchi identity leads to
\ben\label{BI}
{\pa_\nu}{\ln k}\left(\frac{\pa G_k}{\pa \ln k} T_k^{\mu\nu} + G_k\frac{\pa}{\pa \ln k} T_k^{\mu\nu}   \right) = 0.
\een
The existence of a functional relation between the RG scale $k$ and cosmological time $t$ means that \eq{BI} becomes
\begin{equation}\label{dt}
\frac{d\ln k}{d t}\,\frac{\pa (G_k\,\rho_k)}{\pa \ln k}=0\,.
\end{equation} 
From \eq{dt} with $d\ln k/d t\neq 0$ and the Friedmann constraint \eq{Friedmann} it follows that $\pa H/\pa \ln k=0$, reflecting the fact that  $H$ and $k$ are independent phase space variables. Furthermore, using the Friedmann constraint \eq{Friedmann} together with \eq{xyz} for the running couplings \eq{eq:RGexpDef} provides the condition
\ben
\etRG(k) + y^2\, \nuRG(k,z) = 0\,. \label{eq:BiaCon}
\een
Finally, to derive the dynamical equation for the time-dependence of $k$, we differentiate \eq{eq:BiaCon} with respect to time, leading with some algebra to a relation between $d\ln k/dN$ and $x$, $y$, $z$ and $dy/dN$. Using the short-hand notation
\begin{equation} \label{eq:RGalOmDef}
\alRG = \half\left[ \etRG + \nuRG - \frac{\partial}{\partial \ln k} \ln \left(-\frac{\etRG}{\nuRG}\right)\right]\, ,
\end{equation}
we find
\begin{equation}
\frac{d \ln k}{dN}
 =\frac{1}{\alRG}\left[\frac{\siRG}{\nuRG}\sqrt{\frac32}x z + 3 x^2  + \frac{3}{2}\sum_i\ga_i\Om_i  \right]\,.  \label{eq:TeqnRG}
\end{equation}
The significance of this equation is the following.  Under the assumption that the renormalisation scale can be taken to be time-dependent, and that the effect is to introduce time-dependence of the couplings via their dependence on the renormalisation scale,  it follows that $k(N)$ must be chosen to satisfy (\ref{eq:TeqnRG}) in order to maintain the consistency of the evolution equations. The full equations for $x$, $y$, $z$ and $\Om_i$ are to be found by substituting \eq{eq:TeqnRG} into the right-hand sides of \eq{eq:XeqnRG} -- \eq{eq:OeqnRG}.
We stress that the evolution equation for $\ln k$ is not the same as the evolution equation for $\ln H$. The latter remains functionally the same as in the classical set-up  \eq{dHdN}, except that  the right-hand side depends implicitly on the RG parameters.
%, which under our starting assumption outlined at the beginning of Section \ref{RGC} is given by the classical equation of motion \eq{dHdN}. 
Indeed, 
comparing \eq{eq:TeqnRG} and \eq{dHdN}, we note that
\bea
\frac{d \ln H}{dN}
 &=& {\alRG}\, \frac{d \ln k}{dN} -\frac{\siRG}{\nuRG}\sqrt{\frac32}x z\,.  \label{dkdH}
\eea
We can use \eq{dkdH} to read off the necessary conditions  under which $k$ and $H$ evolve the same way with scale factor:
 $\alRG=1$ is mandatory, together with either $\siRG=0$, $1/\nuRG=0$, $x=0$ or $z=0$. We note that $\alRG=1$ holds at a gravitational fixed point. In general, however, we stress that the evolution of $\ln k$ 
 and $\ln H$ 
  with cosmological time is different and hence  $k$  cannot  be chosen to be proportional to $H$ (or $1/t$). Even in the simplified case where $x=0$ or $z=0$, \eq{dkdH} in general does not imply $H\propto k$ because $\alRG$ in \eq{eq:RGalOmDef} has a non-trivial $k$- and $z$-dependence.
\medskip

%In the classical limit, the RG parameters $\etRG$, $\nuRG$, $\siRG$ go to zero. Thus the corrections in Eqs. (\ref{eq:XeqnRG}, \ref{eq:YeqnRG}, \ref{eq:ZeqnRG}, \ref{eq:OeqnRG}) disappear. One has also to require that in the classical limit $\siRG/\nuRG\rightarrow 0$. Then $H$ becomes proportional to $k$. Therefore, $d\ln k/dN$ will not vanish in the classical regime as long as $H$ keeps evolving, but as soon as a de Sitter phase is entered with simultaneously negligible RG effects. During inflation the RG effects in the scalar sector are presumably not negligible. Note in any case that our assumption is that there is no energy exchange between the running couplings and the matter fields. 

In the strict classical limit, all $k$-dependence is removed and the RG parameters $\etRG$, $\nuRG$ and $\siRG$ in \eq{eq:RGexpDef} vanish identically. Consequently, the modified cosmological equations fall back onto their classical counterparts. In addition, the $k$-dependence of \eq{eq:RGexpDef}  vanishes, which entails $\alRG=0$ in \eq{eq:RGalOmDef}. In consequence, the scale-derivative of the Bianchi identity \eq{dt}  no longer provides a constraint for the RG parameters, and the relations \eq{eq:BiaCon} and \eq{dkdH} become empty.  The dynamical equations formally fall back onto the classical ones for vanishing ${d\ln k}/{dN}$. In general this limit does not imply classical behaviour,  because the RG parameters  \eq{eq:RGexpDef} and \eq{eq:RGalOmDef} can be non-zero (see Sec.~\ref{FI}).
One may wonder under which conditions the classical cosmological equations are achieved {\it dynamically}. For generic $k$, this happens provided that
%can ask what happens when the RG parameters $\etRG$, $\nuRG$, $\siRG$ go to zero as $k$ decreases, apparently signalling a classical limit. However, the extra terms in (19-22) become negligible only if 
the RG-induced dependence on cosmological time $\etRG d\ln k / d N$, $\nuRG d\ln k / d N$ and $\siRG d\ln k / d N$ become sufficiently small. The latter terms depend on the ratios $\etRG/\alRG$, $\nuRG/\alRG$,  and $\siRG/\alRG$ due to \eq{eq:TeqnRG}.  
%Hence the equations (19-22) do not necessarily revert to the classical equations (6-9) in this limit. 
Similarly, $k$ continues to evolve according to \eq{eq:TeqnRG} unless the classical parameters $x$, $z$ and $\gamma$ vanish (or $\alRG$ diverges). We conclude that the smallness of  $\etRG$, $\nuRG$, and $\siRG$ alone does not imply  a dynamical approach to the classical equations, although reversion to the classical cosmology is one possibility. If the RG-induced terms remain non-vanishing, this would mean that radiative corrections from Hubble-scale momenta continue to be relevant.

\section{Cosmological fixed points}\label{CFP}

In this section, we discuss cosmological fixed point solutions 
for renormalisation group improved cosmologies 
with one barotropic fluid and a scalar field, and
subject to  the Friedmann constraint 
and  the Bianchi identity. 
Note that we  have six equations in four unknowns, and we should therefore expect to find constraints on the parameters $\etRG$, $\nuRG$, $\siRG$ and $\ga$, and on the form of the function $\et(z)$, for solutions to exist. \medskip
 
To achieve a cosmological fixed point solution in the presence of a scalar field, the renormalisation group scale $\ln k$ may or may not be forced to be constant with scale factor, depending on the values of $x$, $y$, $z$ and $\Om_\ga$, see \eq{eq:TeqnRG}.  Hence two qualitatively different types of cosmological fixed point solutions become available: a fixed point where the RG scale $\ln k$ continues to scale with $N$, and a fixed point where the RG scale becomes independent thereof.\medskip  

In the remainder, we evaluate  the two scenarios in more detail.
We also exhibit five cosmological fixed point solutions in Tab.~\ref{table2}, in four of the categories introduced in Tab.~\ref{table1}.  We are not interested in fluid-dominated solutions as we have assumed that the RG running of the fluid parameters is negligible. \medskip

\subsection{Freeze-in fixed points}\label{FI}
We begin with the scenario where the RG scale parameter $\ln k$ acquires a fixed point under the evolution with cosmological time,
\begin{equation}\label{0}
\frac{d\ln k}{dN}=0\,.
\end{equation} 
The significance of \eq{0} is that the evolution of $\ln k$ with $N$ (or cosmological time) comes to a halt at some freeze-in scale $k_{\rm fi}$. In consequence, the RG parameters stop to evolve. The cosmological fixed points \eq{FP} are formally the same as those of the classical set-up, and clearly the classical fixed points appear as a particular solution of \eq{0} where no RG parameter dependence enters from the outset.  Here, however, quantum corrections are present and incorporated through  RG modifications in the variables $x$, $y$ and $z$.\medskip

At freeze-in, we learn from  \eq{dkdH}  that $d\ln H/dN$ can remain non-zero even though the RG scale no longer varies with cosmological time \eq{0}. Also, using \eq{dHdN} and \eq{dkdH} we note that \eq{0} entails a relation between $z$, $x$, $\Omega_i$, and $\nuRG/\siRG$ at the freeze-in scale. For example, at $x\neq 0$, and in the absence of a barotropic fluid, the relation reads  $z=-\sqrt{6}\, x\, \nuRG/\siRG$. Most interestingly, the RG couplings can take generic values at freeze-in, different from the RG fixed points of the underlying quantum field theory.\medskip 

\begin{table*}[t!]
\begin{center}
\begin{tabular}{|c|c|c|c|c|c|c|c|} \hline
{\multirow{3}{*}{Case}}
&{\multirow{3}{*}{$x$}} 
&{\multirow{3}{*}{$y$}}
 & {\multirow{3}{*}{$z$}}
 & {\multirow{3}{*}{$\Omega_{\gamma}$}} 
 &{\multirow{3}{*}{ $\displaystyle \frac{d \ln k}{dN}$}} 
 &{\multirow{3}{*}{ existence}} 
&{\multirow{3}{*}{type}}
\\&&&&&&&
\\&&&&&&&
\\ \hline
{\multirow{3}{*}{(a)}}
&{\multirow{3}{*}{ 0}}
 &  {\multirow{3}{*}{$1$ }}
 &{\multirow{3}{*}{ 0}}
& {\multirow{3}{*}{ 0}}
 &{\multirow{3}{*}{ - }} 
 & {\multirow{3}{*}{$\displaystyle -\frac{\etRG}{\nuRG}=1$}} 
  &{\multirow{3}{*}{potential}}
  \\
  &&&&&&&
  \\
  &&&&&&&
\\ \hline
{\multirow{3}{*}{(b)}}
&{\multirow{3}{*}{  $ \pm 1$ }}
&{\multirow{3}{*}{  0}}
 &{\multirow{3}{*}{ $z_*$}}
  &{\multirow{3}{*}{ 0}} 
  &{\multirow{3}{*}{ $\displaystyle  \frac{6}{\nuRG}\left(1 \pm\left( \frac{\siRG}{\nuRG}\right)\frac{z_*}{\sqrt{6}} \right)$ }}
   &{\multirow{3}{*}{$\displaystyle {\etRG}=0$}}
 &{\multirow{3}{*}{kinetic}}
 \\
 &&&&&&&
 \\
  &&&&&&&
\\ \hline
{\multirow{3}{*}{(c1)}}
& {\multirow{3}{*}{   $\displaystyle \pm\left(1 + \frac{\etRG}{\nuRG} \right)^\half$ }}
&{\multirow{3}{*}{  $\displaystyle \left(- \frac{\etRG}{\nuRG} \right)^\half$}} 
 &{\multirow{3}{*}{ $0$}}
 & {\multirow{3}{*}{0}}
 &{\multirow{3}{*}{$\displaystyle \frac{6}{\nuRG}$}} 
 &{\multirow{3}{*}{ $\eta(0) = 0$}} 
&{\multirow{3}{*}{mixed}}
\\
&&&&&&&
\\
 &&&&&&&
   \\ \hline
{\multirow{3}{*}{(c2)}}
& {\multirow{3}{*}{   $\displaystyle \pm\left(1 + \frac{\etRG}{\nuRG} \right)^\half$ }}
&{\multirow{3}{*}{  $\displaystyle \left(- \frac{\etRG}{\nuRG} \right)^\half$}} 
 &{\multirow{3}{*}{ $z_*$}}
 & {\multirow{3}{*}{0}}
 &{\multirow{3}{*}{$\displaystyle \frac{6}{\nuRG}\left( 1+ \frac{z_*}{\sqrt{6}x_*} \right)$}} 
 &{\multirow{3}{*}{ $\siRG = \nuRG$}} 
&{\multirow{3}{*}{mixed}}
\\
&&&&&&&
\\
 &&&&&&&
   \\ \hline
{\multirow{3}{*}{(d1)}}&
{\multirow{3}{*}{  $0$ }}
&{\multirow{3}{*}{ $\displaystyle   \left(- \frac{\etRG}{\nuRG} \right)^\half$}}
 &
 {\multirow{3}{*}{ $0$}}
 &{\multirow{3}{*}{ $\displaystyle 1 + \frac{\etRG}{\nuRG}$}}
   &{\multirow{3}{*}{ $\displaystyle \frac{3\gamma}{\nuRG}$ }}
&{\multirow{3}{*}{$\gamma\neq 0$ }}  
 &{\multirow{3}{*}{scaling}}
 \\
&&&&&&&
 \\
 &&&&&&&
\\
\hline
{\multirow{3}{*}{(d2)}}&
{\multirow{3}{*}{  $\displaystyle \pm\left(-\frac{\etRG}{\nuRG}\frac{\ga}{2-\ga}\right)^\half$ }}
&{\multirow{3}{*}{ $\displaystyle   \left(- \frac{\etRG}{\nuRG} \right)^\half$}}
 &
 {\multirow{3}{*}{ $\displaystyle -\sqrt{\frac{3}{2}}\frac{\ga}{x_*}$}}
 &{\multirow{3}{*}{ $\displaystyle 1 + \frac{\etRG}{\nuRG}\frac{2}{2-\ga}$}}
   &{\multirow{3}{*}{ $0$ }}
&{\multirow{3}{*}{  $\siRG = \nuRG$}}  
 &{\multirow{3}{*}{scaling}}
 \\
&&&&&&&
 \\
 &&&&&&&
\\
\hline
\end{tabular}
\caption{\label{table2} {Renormalisation group improved cosmological fixed points for Einstein gravity with a scalar field
 and one other barotropic fluid.
 We adopt the classification of Tab.~\ref{table1}. The RG parameters are defined in \eq{eq:RGexpDef}.  In case (a), the conditions for $d\ln k/dN$ are discussed in Sec.~\ref{pot}. 
In case (b), $z_*$ is a solution to \eq{eq:zEqKinDom}, and in case (c2) $z_*$ is a solution to \eq{eq:zEqnMix} (see text).
}}
\end{center}
\end{table*}

\subsection{Simultaneous fixed points}
A second type of cosmological fixed points is achieved for 
\begin{equation}\label{const}
\frac{d\ln k}{dN}={\rm const.}\neq 0\,.
\end{equation}
Here, the RG scale continues to evolve with cosmological time. The full system depends on $k$ implicitly through the variables \eq{xyz}, and explicitly through the RG parameters \eq{eq:RGexpDef}. Hence, a cosmological fixed point 
with \eq{const} can only be achieved if the RG parameters \eq{eq:RGexpDef} have become $k$-independent. 
This includes, in particular, a RG fixed point of the underlying quantum field  theory, in which case the coefficient \eq{eq:RGalOmDef} simplies and reads $\alRG = \half(\etRG+\nuRG)$. 
\medskip

\subsection{Potential-dominated fixed point}  \label{pot}
In Tab.~\ref{table2}~(a), the potential-dominated fixed point is an inflating one, which requires that the RG parameters are related by $\etRG = -\nuRG$.  One way of realising this, together with \eq{const}  is to have an asymptotically safe fixed point for the gravitational coupling, at which $\etRG = -2$, and an RG scaling fixed point for the scalar field, with the field $\phi$ approaching infinity.  Writing the potential as $V(\phi) = \la_2\phi^2$, we see that  $z = 2/(\ka \phi)$, and hence the cosmological fixed point is at $\phi\to\infty$. We also see that $\nuRG = \siRG = \be_{2}/\la_2$, where $\be_2$ is the $\be$-function for $\la_2$ (twice the mass squared).  To obtain $\nuRG = 2$, we require that $\la \to k^2\tilde\la_2$, where $\tilde\la_2$ is a dimensionless constant.  Although a recent RG investigation of a sub-class of theories of scalar fields coupled to gravity \cite{Narain:2009fy,Narain:2009gb} showed no sign of such a fixed point in four dimensions, this does not rule one out altogether. The consequences of such a fixed point are very interesting and we will return to them in the concluding section.  Note that one needs the form of $\etRG(k)$ and $\nuRG(k)$ in order to be able to evaluate $d \ln k/dN$ at the fixed point.\medskip

Alternatively, another inflating fixed point is realised for general potentials through the freeze-in scenario \eq{0} which entails $d\ln H/dN=0$. This  still requires $\nuRG = -\etRG$, though together with $\alRG\neq 0$.  For this scenario to be applicable we note that $\frac{\partial}{\partial \ln k} \ln (-\etRG/\nuRG)$ cannot vanish at freeze-in, meaning that $\nuRG$ and $\etRG$ cannot simultaneously achieve fixed points in the underlying field theory.

\subsection{Kinetic-dominated fixed point}  
A kinetic-dominated fixed point  requires the vanishing of the graviton anomalous dimension $\etRG \to 0$ together with $\nuRG$ non-zero, see  Tab.~\ref{table2}~(b). The first condition implies that a kinetic-dominated fixed point can only be achieved when gravity is essentially classical near the Gaussian fixed point of Einstein gravity,  where $G_k$ and $\La_k$ are constant in the infrared. In addition, for \eq{const}, $z$ must be a solution to the equation
\ben\label{eq:zEqKinDom}
\et(z) = z^2\left(1- \frac{\siRG}{\nuRG}+\left(\frac{\siRG}{\nuRG}\right)^2 \right)\mp \sqrt{6} \left(1- \frac{\siRG}{\nuRG} \right)z \ ,
\een
with the sign depending on whether $x_* = \pm 1$.  This can be achieved either with an exponential potential $V = V_0\exp(\la\ka\phi)$ with $\la = \mp \sqrt 6/\left({\siRG}/{\nuRG}+1\right)$, or a potential with the intriguing form
\ben\label{eq:Vintrigue}
V = V_0\left(1 - \exp{\la Q\ka \phi}\right)^{-{1}/{Q}}.
\een
where $Q = \left({\siRG}/{\nuRG}\right)^2-1$. Alternatively, the freeze-in scenario  \eq{0} is realised for general potentials provided that $z$ settles to either of the values $z=\mp (2\nuRG)/(\sqrt{6}\siRG)$ at the freeze-in scale.

\subsection{Mixed fixed points}  

In fixed points with both kinetic and potential energy,  but no barotropic fluid component, one finds that the $dx/dN = 0$ equation reduces to 
\ben
z\,y^2\,\left( 1 - \frac{\siRG}{\nuRG}\right) = 0\ ,
\een
see  Tab.~\ref{table2} (c). Consequently, we either have $z=0$ (c1), or $\nuRG = \siRG$ (c2). Case (c1) can be achieved with a monomial potential $\la_n\phi^n$ and $\phi \to \infty$. If the theory approaches an asymptotically safe Gaussian matter fixed point \cite{Percacci:2003jz,Narain:2009fy,Narain:2009gb} where all couplings of the scalar vanish and $V \to \la_0$, we have $\etRG \to -2$, $\nuRG \to 4$, and there is equipartition between kinetic and potential energy: $x^2 = y^2 = \frac 12$. In case (c2) stationarity of $z$ requires that it must be a solution to the equation
\ben\label{eq:zEqnMix}
-\et(z) + z^2\left( 1 +  R \right) + \sqrt{6} x_*\, R\,  z = 0\ ,
\een
where $R = y_*^2/(2x_*^2)$.  This can be achieved in two ways, either with a potential with same form as \eq{eq:Vintrigue}, but with $\la = \sqrt{6} x_*$ and $Q =  R$ and \eq{const}, or as a freeze-in scenario \eq{0}  provided that $z_*=-\sqrt{6}x_*$ and thus $\eta=z^2$, together with  $\nuRG = \siRG$, leading to an exponential potential $V = V_0\exp(-\sqrt{6}x_*\ka\phi)$.  Provided
\begin{equation}
\frac{\etRG}{\nuRG} < -\frac23\,,
\end{equation}
the expansion is accelerating at these fixed points.

\subsection{Scaling fixed point}

A scaling fixed point  is found for either \eq{const} corresponding to case (d1), or during freeze-in corresponding to case (d2). The case (d1) requires $\gamma\neq 0$, and the fixed point values for $y$ and $\Omega$ are then uniquely fixed by RG parameters. Interestingly, a special case for the fixed point  (d1) has been found previously for a universe with a cosmological constant and a barotropic fluid (no scalar field), provided that gravity approaches an asymptotically safe RG fixed point with $\Om_\La = \frac 12$ ~\cite{Bonanno:2001xi,Bonanno:2001hi,Bentivegna:2003rr,Reuter:2005kb}.  In our conventions, a cosmological constant $\Lambda_k$ is equivalent to a stationary scalar field with $x=0$ and potential $V_k = \La_k/(8\pi G_k)$, together with \eq{CC}. If gravity becomes asymptotically safe,  we have $\etRG = -2$, $\nuRG= 4$ and $\alRG=1$ (see Sec.~\ref{RG}). Therefore  the Bianchi identiy leads to $y^2 \equiv \Om_\La = \frac 12$. Hence, the field-theoretical  RG fixed point of asymptotically safe Einstein-Hilbert gravity becomes enhanced into a cosmological fixed point under cosmological time evolution. Comparing \eq{eq:TeqnRG} with \eq{dHdN} for $x=0$ and $\alRG=1$ we also confirm that $d\ln k/dN=d\ln H/dN$, which explains the scaling relation  $H\propto k$ at this fixed point. \medskip

If one looks for a trial solution with \eq{0}, the same argument for the mixed cosmological fixed points above applied to the case where $\Om_\ga \ne 0$ leads again to the requirement that $\nuRG = \siRG$ at freeze-in,  see Tab.~\ref{table2} (d2).  For such a solution we require that $\et(z) = z^2$, and $z_* = -(\sqrt{3}\ga)/(\sqrt{2}x_*)$. This means that we must have an exponential potential with 
\ben
\la =  -\sqrt{\frac{3\ga(2-\ga)}{2}\left(-\frac{\nuRG}{\etRG}\right)}\ .
\een
It is somewhat peculiar that the potential parameter is related to the fluid equation of state and the RG parameters in this way, and it would seem hard to achieve in practice.  Note that this solution is an accelerating one provided that $\ga<\frac23$, which is the same as the classical condition. 

\section{Conclusions}

In this paper we have examined the application of renormalisation group 
ideas to scalar field cosmology.
We have shown how to relate  
RG scale evolution to cosmological evolution in such a way as to maintain the form of the 
classical equations of motion. 
Preserving the Bianchi identity during the cosmological evolution means that 
the RG scale   must obey \eq{eq:TeqnRG}: one cannot simply assume that it is (for example) proportional to 
the Hubble parameter $H$ from the outset.\medskip

Given this consistency relation we have written down the cosmological 
evolution equations with running gravitaitonal and matter couplings
\eq{eq:XeqnRG} - \eq{eq:OeqnRG}, showing that the RG scale dependence
is transmitted via the quantities $\etRG$, $\nuRG$, and $\siRG$ 
\eq{eq:RGexpDef}.
We have identified the fixed points of the cosmological evolution for a 
universe whose energy-momentum receives contributions from a scalar field and 
a single barotropic fluid (Tab.~\ref{table2}), showing under what conditions they 
appear, and relating them to fixed points in the classical case (Tab.~\ref{table1}).\medskip

At the fixed points the Universe may or may not accelerate,  depending on the ratio of the RG parameters $\etRG$ and $\nuRG$. 
An inflating cosmology controlled by an asymptotically safe fixed point  where $\etRG = -2$ requires that 
$\nuRG < 3$. In particular a Gaussian matter fixed point ($\nuRG = 4$)
 does not produce accelerated expansion. 
A fixed point resembling standard slow-roll inflation ($x \to 0$) requires $\nuRG \to 2$, which could be achieved by a scalar field with a quadratic potential whose mass parameter runs proportional to the RG scale. Alternatively, the RG scale can freeze, leading to non-universal values for the parameters $\etRG$ and $\nuRG$. 
Other scalar field dominated fixed points exist, where the energy density is divided between kinetic and potential term of the scalar field. In the case of the Gaussian matter fixed point, this distribution happens in equal parts. There is also a kinetic-dominated fixed point and a scaling fixed point. 
\medskip

A crucial question which we leave for future publication is how the universe evolves towards or away from 
fixed points.  To answer the question one must have a specific scalar potential in mind, and explicit forms for the couplings as a function of the RG scale.  One can then use the equations we have developed to follow the universe near the fixed point, and study the transition to or from classical FRW evolution.
Exactly at what scale this transition takes place is not clear, or whether the RG fixed point is departed before or after the cosmological one if they are simultaneous. 
A conservative view would be that the transition scale is the Planck scale, well above the 
Hubble scale during inflation, and so observational effects on inflationary perturbations are presumably 
small. Departures from classical slow-roll inflation can be probed using the 
increasing amount of detailed information about the primordial 
fluctuations, and so constraints on the transition scale can be derived from observations.
Likewise, scenarios involving a universe with a 
scalar field approaching the more speculative IR fixed point can be 
tested against cosmological data such as Type Ia supernovae \cite{Kowalski:2008ez,Bentivegna:2003rr,EspanaBonet:2003vk}.\\[2ex]

\noindent{\bf Acknowledgements}\ \ This work was supported by the Science and Technology Research Council [grant number ST/G000573/1].

\bibliography{AS_references}

\end{document}